\documentclass[12pt]{article}
\usepackage{amsmath,amssymb,graphicx,mathrsfs}
\usepackage{hyperref}
\newcommand{\be}{\begin{equation}}
\newcommand{\ee}{\end{equation}}
\newcommand{\bea}{\begin{eqnarray}}
\newcommand{\eea}{\end{eqnarray}}

\def\({\left(} \def\){\right)}

\def\revise#1       {\raisebox{-0em}{\rule{3pt}{1em}}
                     \marginpar{\raisebox{.5em}{\vrule width3pt\
                     \vrule width0pt height 0pt depth0.5em
                     \hbox to 0cm{\hspace{0cm}{%
                     \parbox[t]{4em}{\raggedright\footnotesize{#1}}}\hss}}}}

\begin{document}
\title{The ratio of  shear viscosity to entropy density in generalized theories of gravity}
\author{Ram Brustein \\ Department of Physics, Ben-Gurion University,\\
    Beer-Sheva 84105, Israel \\ E-mail: ramyb@bgu.ac.il\\ \\
A.J.M. Medved \\ Physics Department, University of Seoul, \\
Seoul 130-743 Korea \\
    E-mail: allan@physics.uos.ac.kr}

\maketitle


\abstract{
Near the horizon of a black brane solution in Anti-de Sitter space, the long-wavelength fluctuations of the metric exhibit hydrodynamic behaviour. For Einstein's theory, the ratio of the shear viscosity of near-horizon metric fluctuations $\eta$ to the entropy per unit of transverse volume $s$ is  $\eta/s=1/4 \pi$. We propose that, in generalized theories of gravity,  this ratio is given by the ratio of two effective gravitational couplings and can be different than $1/4 \pi$.
Our proposal implies that $\eta/s$ is equal for any pair of gravity theories that can be transformed into each other by a field redefinition. In particular,
the ratio  is $1/4\pi$ for any theory  that can be transformed into Einstein's theory; such as $F(R)$ gravity.
Our proposal also implies that matter interactions
--- except those including explicit or implicit
factors of the Riemann tensor --- will not modify $\eta/s$.
The proposed formula reproduces, in a very simple manner, some recently found results for Gauss-Bonnet gravity. We  also make a prediction for $\eta/s$ in Lovelock theories of any order or dimensionality.
}



\newpage
\label{intro}

The translation invariance of the horizon of a black brane in Anti-de Sitter (AdS) space implies that the long-wavelength fluctuations of the near-horizon metric  exhibit hydrodynamic behaviour  and satisfy a Kubo formula from which the shear viscosity coefficient $\eta$ can be extracted \cite{hydroI}. The near-horizon hydrodynamics seems to be related to the hydrodynamics of the metric fluctuations near the boundary of AdS space, which in turn can be related via the AdS/CFT duality to the hydrodynamics of strongly coupled gauge theories \cite{hydroIII}. The latter provides an interesting theoretical framework for studying relativistic hydrodynamics and may explain the experimental results of
heavy-ion collisions.

Black branes are known to possess an entropy \cite{bekenstein,hawking} proportional to the area of their horizon.
For black brane solutions of Einstein's gravity, the entropy per unit of transverse volume $s$ is equal to a quarter of the horizon area in units of Newton's constant. A focus of attention in this context  has been
on the ratio $\eta/s$.  For Einstein's gravity,  this value is well known: $\eta/s=1/4 \pi$ \cite{PSS1}. Recently,
it has become clear that the ratio $\eta/s$ can be corrected in the presence of
higher-derivative gravity corrections \cite{myers,buchel}.
Such interactions are important because they may allow a better calculation of this ratio in strongly coupled gauge theories. The general theories of gravity that are relevant to this duality can be viewed as perturbative corrections to Einstein's gravity.

We propose a general formula for calculating the ratio $\eta/s$ for any theory of gravity. Our idea is based on a recent reinterpretation \cite{BGH} of the Wald formula \cite{wald1,wald2} for the dynamical entropy of black branes in general theories of gravity.  Our formula is local: It only requires  knowing the corrected Lagrangian and its value on the horizon of the black brane.

Let us consider perturbations of the brane metric  $g_{\mu\nu}\rightarrow g_{\mu\nu} + h_{\mu\nu}$
and single out $z$ as  the propagating direction of a graviton
on the brane. Under a suitable choice of gauge \cite{PSS2}, it is found that the highest-helicity
polarization for the $z$-propagating perturbations $h_{xy}$
decouple from all others ---  making this class of gravitons particularly convenient for deducing
hydrodynamic parameters.

Momentarily restricting to Einstein's gravity, one obtains the action
$
I = -\frac{1}{32 \pi \kappa_E^2}\int dr dt d^dx \sqrt{-g}\frac{1}{2}g^{\mu\nu}\partial_\mu
\phi \partial_\nu\phi +\cdots \;
\label{action}
$
for the  field $\phi\equiv h^{x}_{\ y}$.
(Here, $\kappa_E^2=G_N/2$ is the gravitational coupling  in terms of Newton's constant
and the dots denote higher-order terms.)
The situation does not change by much for a general theory of gravity.
The action will maintain its basic form, but now the gravitational coupling will be  generally different from $\kappa^2_E$ and can be polarization dependent (as elaborated on below).

\label{theratio}

Our task is, for a general theory of gravity, to calculate the shear viscosity $\eta$ near the horizon of a black brane in terms of the effective coupling of the field $\phi=h^x_{\ y}$. Let us denote this specific
coupling as $\kappa^2_{xy}$ (and, in general, denote couplings by $\kappa^2_{\mu\nu}$).

The  proposed calculation  can be carried out in three steps:
First of all, it was pertinently demonstrated in \cite{KSS} that
the field equation for $\phi$ is equivalent to that of a minimally coupled scalar
field. One might be concerned that the calculation of \cite{KSS} was specific to Einstein's gravity.
However,  recall our assertion that a general theory can be modeled perturbatively (at the two-derivative level) as Einstein's gravity with a modified gravitational coupling. Because hydrodynamics entails the stationary (zero-momentum)
limit of the gravitons in the neighborhood of the horizon, any such
coupling can reasonably be regarded as a constant parameter.
This realization is sufficient to ensure that the generalized calculation will carry through unfettered.

Secondly, the identification of $\phi$ with a scalar is significant for the following reason:
The absorption cross-section at zero momentum of a minimally
coupled scalar by  a black brane
has been previously calculated \cite{DGM,emparan}. The result for
this cross-section $\sigma$ is, in fact, remarkably simple:
$
\sigma=\textit{\Large a}
$,
with $\textit{\Large a}$ denoting the horizon area per unit of transverse volume.

Thirdly, the prior outcome has further relevance because of a well-established connection between
the shear viscosity and the absorption cross-section of a black brane.
For asymptotically flat spacetimes, arguments of a general nature have lead to the deduction
that $\eta\propto\sigma$ \cite{klebanov,gubser}, more specifically, $\eta=\frac{\sigma}{32\pi \kappa_{E}^2}$. This relation has since,
with the help of the Kubo-formula machinery,
been made very precise \cite{PSS1,Benincasa:2006fu}. Our result for generalized theories of gravity then follows:
\begin{equation}
\eta= \frac{\textit{\Large a}}{32\pi \kappa_{xy}^2}\;,
\end{equation}
where we have formally replaced  $\kappa^2_E$ with $\kappa^2_{xy}$.
The gravitational coupling $\kappa^2_{xy}$ need not be as simple as $G_N/2$ nor be independent of the polarization. This distinction is very important in what follows.

Let us now reconsider the entropy density $s$. For Einstein's  theory,
$s=\textit{\Large a}/4G_N=\textit{\Large a}/8\kappa_E^2$. So then what happens for a general gravity theory?
As explained in \cite{BGH}, Wald's Noether-charge formalism \cite{wald1,wald2}
uniquely picks out the $h_{rt}$ gravitons as the class that is relevant to  entropy,
\begin{equation}
s=\frac{\textit{\Large a}}{8 \kappa_{rt}^2}  \;.
\end{equation}

Combining the last two equations, we obtain
\begin{equation}
\frac{\eta}{s}=\frac{1}{4\pi}\ \frac{ \left(\kappa_{rt}\right)^2}{\left(\kappa_{xy}\right)^2}\;.
\label{ratio}
\end{equation}
This is our proposed formula for the viscosity---entropy ratio.
Obviously, this reduces to the standard result  of $1/4\pi$ if the gravitational
couplings are equal. But, as will be shown in the examples to follow,
this need not necessarily be the case. The couplings can be corrected by higher-order gravitational corrections or by complicated-enough matter corrections.

The gravitational coupling for a specific polarization was defined
in \cite{BGH},  where further details are given. We simply present the needed result:
\begin{equation}
\label{kappadefgeneral}
\frac{1}{\left(\kappa_{\mu\nu}\right)^2}=\mp\frac{1}{4}
\left( \frac{\delta\mathscr{L}}{\delta R_{ab}^{~~cd}}\right)^{\!\!(0)} \hat\epsilon_{ab}\hat\epsilon^{cd}\;,\;\;\;
\{a,b,c,d\}\in\{\mu,\ \nu\}\;.
\end{equation}
Here, $\mathscr{L}$ is the Lagrangian of interest and
$\hat\epsilon_{ab}$ is the binormal vector with regard to the specified pair of polarization directions.
Any binormal is antisymmetric under the exchange of $a$ and $b$,
and normalized such that $\hat\epsilon_{ab}\hat\epsilon^{ab}=\mp 2$. A $\mp$ sign is only to be taken as negative when one of the directions $(\mu,\nu)$
is timelike; with this distinction ensuring the positivity of the coupling.
The superscript $(0)$ signifies that the calculation is always made on solution {\em and} on the horizon.

Although our proposal is valid for any theory of gravity, for future reference  it is convenient to reformulate the above  in terms of a manifestly Einstein-corrected  theory.  Let us suppose a Lagrangian of the form $\mathscr{L}=\frac{1}{32\pi\kappa_E^2}\left[R+
\lambda \mathscr{L}_C\right]$, where
$\lambda$ is a ``perturbative" parameter and $\mathscr{L}_C$  represents the
correction.
It can then be shown that
\begin{equation}
\label{kappadefpert}
\frac{\left(\kappa_E\right)^2}{\left(\kappa_{\mu\nu}\right)^2}=
1\mp\frac{\lambda}{2}
\left( \frac{\delta\mathscr{L}_C}{\delta R_{ab}^{~~cd}}\right)^{\!\!(0)} \hat\epsilon_{ab}\hat\epsilon^{cd}\;,\;\;\;
\{a,b,c,d\}\in\{\mu,\nu\}\;,
\end{equation}
where it is sufficient to use the lowest-order (Einstein) solution to obtain
the result to  the leading order in $\lambda$.

\label{examples}

Let us now discuss some explicit examples for which a definitive statement about the ratio
$\eta/s$ can be made. We  present these examples to study the implications of our formula, explicitly demonstrate its simple implementation and provide a preliminary verification of its correctness.


Many ``exotic" gravity theories can be directly transformed into
Einstein's gravity by way of a Weyl transformation or some other field
redefinition \cite{kang}. In this case, one ends up obtaining an uncorrected Einstein theory
albeit with a modified gravitational coupling.
Nonetheless, the revised coupling will necessarily be independent of the polarization.
And, since there are no further deviations from Einstein's theory, it follows that $\eta/s=1/4\pi$
[{\it cf}, eq.(\ref{ratio})].
Interesting examples of this type are Lagrangians with a correction of the form $F(R)$, where
$F$ is an arbitrary function of the scalar curvature. Additionally, those with a correction of the form
$\alpha R^2 +\beta R_{ab}R^{cd}$, with $\alpha$ and
$\beta$ being numerical coefficients. For this very last example,  the
same point was already made in \cite{myers}.

We can advance this basic logic one step further and make the following observation: Any two theories that are related to each other by a field redefinition will have precisely the same value for the
viscosity--entropy ratio. Generally, this ratio will {\em not} be $1/4\pi$.


It is clear from our formalism that $\eta/s$ is solely determined by gravitational couplings at the horizon.
Hence, the addition of matter sources and interactions that do not include --- explicitly or implicitly --- the Riemann tensor can have no effect on the viscosity--entropy ratio.
The way that a matter coupling can depend implicitly on the Riemann tensor is,
as explained in \cite{wald2},
through any of the covariant derivatives acting upon the matter fields. Such couplings do occur
at high-enough orders in string theory.
Meanwhile, it has been observed that some simpler types of matter
do not modify the viscosity--entropy ratio
({\it e.g.}, \cite{mas,son,saremi,maeda,Mateos:2006yd,Benincasa:2006fu,cai}):
Our proposal indicates that this pattern
is really a universal feature of black brane hydrodynamics.


For the following examples, we consider more specifically a black $D-2$  brane in  a
$D$-dimensional AdS spacetime (with $D\geq 5$).
Let us denote the transverse directions as $x_i, \ i=1,\ldots, D-3$, and $z$.
For the solution that we are interested in, the metric $g_{\mu\nu}$ has the following form:
$
ds^2=-f(r)dt^2+\frac{dr^2}{f(r)}+\frac{r^2}{L^2}\left(\sum_idx_i^2+dz^2\right)
$,
where $L$ is the curvature radius of AdS space, and
the function $f(r)$  will depend on the particular theory of gravity  but will
always vanish at the horizon of the black brane.

Let us discuss a correction to Einstein's gravity of the
form $\lambda \mathscr{L}_C =
\lambda R_{\mu\nu\sigma\rho}R^{\mu\nu\sigma\rho}$. This case provides an
interesting check of our proposal, as the  resultant correction to $\eta/s$
has already been  computed explicitly in \cite{myers}.
Since we are treating the problem perturbatively,
it is sufficient to work with the Einstein solution for which
$
f={r^2\over L^2}\left[1-\frac{r_+^{D-1}}{r^{D-1}}\right]
$,
where  $r_+$ is the horizon radius.

It is a straightforward exercise
to compute  the  relevant Riemann-tensor components. Given that
the polarization pairings of interest are $x-y$ and $r-t$, the required
components are summarized by the following pair:
$
R^{tr}_{~~tr}=-\frac{1}{2}\partial_r^2f\rightarrow \frac{(D-1)(D-4)}{2L^2}\;,
$
$
R^{xy}_{~~xy}=-\frac{f}{r^2}\rightarrow 0
$,
with the right-most expressions representing the respective horizon values.

It is trivially clear that the viscosity coupling $\kappa^2_{xy}$ receives no contribution
from the Riemann-squared term, so that
$
\left(\kappa_{xy}\right)^2=\left(\kappa_E\right)^2
$.

To complete the calculation for $\kappa^2_{rt}$, let us make note of the following observation:
Contracting the curvature component $R^{ab}_{~~cd}$ with the binormal vectors $\hat\epsilon_{ab}\hat\epsilon^{cd}$ (where $\{a,b,c,d\}=\{r,t\}$) and  judiciously applying
the (anti-) symmetry properties of the Riemann tensor, one readily obtains $-4 R^{tr}_{~~tr}$.
Yet another factor of 2 comes from varying  $R_{\mu\nu\sigma\rho}R^{\mu\nu\sigma\rho}$.
Incorporating all this into eq.(\ref{kappadefpert}), we then have
\begin{equation}
\frac{1}{\left(\kappa_{rt}\right)^2}= \frac{1}{\left(\kappa_E\right)^2}
\left[1+\frac{2\lambda}{L^2}(D-1)(D-4)\right]\;.
\end{equation}

So it follows that the corrected viscosity--entropy ratio, to leading order in $\lambda$, becomes
\begin{equation}
\frac{\eta}{s}
=\frac{1}{4\pi}\frac{ \left(\kappa_{rt}\right)^2}{\left(\kappa_{xy}\right)^2}
= \frac{1}{4\pi}\left[1-\frac{2\lambda}{L^2}(D-1)(D-4)\right] \;.
\end{equation}
It is easy to check that this result fully agrees with that
found previously in \cite{myers}.


Another instructive example is the Gauss-Bonnet theory, which has  a correction of the form
$
\lambda \mathscr{L}_C =
\lambda \left[R^2 -4 R_{\mu\nu}R^{\mu\nu} + R_{\mu\nu\sigma\rho}R^{\mu\nu\sigma\rho}\right]
$.
Since Gauss-Bonnet gravity and the preceding Riemann-squared theory
are related by a field  redefinition, we already know that the two theories
must necessarily yield the same value for $\eta/s$.
It would, however, be nice to test this point directly.

A ``brute-force"  type of calculation would be one viable way to proceed. But
we will, rather, choose a somewhat different route that is based on
the following identification: The Gauss-Bonnet Lagrangian
can also be regarded as the second-order term in the expansion of
a Lovelock Lagrangian. To elaborate, Lovelock gravity can be viewed as
the most general form of  (gravitational) Lagrangian for which the  field equations will contain no more than two derivatives.
In this sense, the Lovelock theory can be regarded as a natural way of generalizing Einstein
gravity. The Lovelock Lagrangian can  be expressed (schematically) as follows:
$
\mathscr{L}_{LL}=\frac{1}{32\pi\kappa^2_E}\sum_{m=0}^{[D/2]} \lambda_m\mathscr{L}_m\;,
$
where $[...]$ represents the largest integer and $\lambda_m$ is a theory-specific constant
of dimension length$^{2(m-1)}$. Without getting into the intricate details (which can be found,
for instance, in \cite{lovelock}), let us point out that $\mathscr{L}_{0}=1$
represents a cosmological constant, $\mathscr{L}_{1}=R$ is Einstein's gravity,
$\mathscr{L}_{2}$ reduces to the Gauss-Bonnet Lagrangian  and the higher-order terms become
progressively more baroque.
Note that any term will identically vanish, on solution, once $m$ exceeds $D/2$.

This identification is helpful because of an outcome that was derived some time ago \cite{ted,visser}:
$
\left( \frac{\delta\mathscr{L}_m(g)}{\delta R_{ab}^{~~cd}}\right)^{\!\!(0)} \hat\epsilon_{ab}\hat\epsilon^{cd}
= -2m\left(\mathscr{L}_{m-1}(g_{\parallel})\right)^{\!\!(0)}$, where
$\{a,b,c,d\}\in\{r,t\}
$.
Some clarification of the symbolism is certainly in order. Firstly, $\mathscr{L}_{p}$
denotes the $p$-th order term of the Lovelock expansion.
Meanwhile,  the ``transverse" metric  $g_{\parallel}$ has a very special meaning:
In calculating the lower-order Lagrangian, one is instructed to still use the actual metric $g$ but to disregard all
Riemann-tensor components carrying one or more  $r$ and/or $t$ indices.

Retracing the steps of \cite{ted,visser}, one should be convinced
that the converse is also true. That is,
$
\left( \frac{\delta\mathscr{L}_m(g)}{\delta R_{ab}^{~~cd}}\right)^{\!\!(0)} \hat\epsilon_{ab}\hat\epsilon^{cd}
= -2m\left(\mathscr{L}_{m-1}(g_{\perp})\right)^{\!\!(0)}$, with
$ \{a,b,c,d\}\in\{x,y\}
$,
where the ``normal" metric $g_{\perp}$ advises one to  disregard all Riemann-tensor components
carrying one or more transverse (brane) indices.

Given that  $\mathscr{L}_2$ represents  Gauss-Bonnet gravity and $\mathscr{L}_1$ is simply
Einstein's gravity, the calculation can now proceed in a straightforward manner.
Since all curvature components of the form $R^{xy}_{~~xy}$ have been shown to  vanish trivially
on the horizon, we must have  that $\mathscr{L}_1(g_{\parallel})=0$, and so
$
\left(\kappa_{rt}\right)^2=\left(\kappa_E\right)^2
$.
Whereas, following the above prescription, we can further deduce that
\begin{equation}
\mathscr{L}_1(g_{\perp})=R(g_{\perp})=R^{tr}_{~~tr}+R^{rt}_{~~rt}=-\partial_r^2 f\rightarrow
\frac{(D-1)(D-4)}{L^2}\;.
\end{equation}
Inserting this result into  the $x-y$ version of eq.(\ref{kappadefpert})
and also taking into consideration the additional factor of $-2m=-4$,
we then obtain
\begin{equation}
\frac{1}{\left(\kappa_{xy}\right)^2}= \frac{1}{\left(\kappa_E\right)^2}
\left[1-\frac{2\lambda}{L^2}(D-1)(D-4)\right]\;.
\end{equation}
Consequently,
\begin{equation}
\label{etasGB}
\frac{\eta}{s}=\frac{1}{4\pi}\frac{ \left(\kappa_{rt}\right)^2}{\left(\kappa_{xy}\right)^2}
= \frac{1}{4\pi}\left[1-\frac{2\lambda}{L^2}(D-1)(D-4)\right] \;.
\end{equation}
We have, as anticipated, duplicated our previous finding for Riemann-squared (corrected) gravity.

In spite of this confirmation, the reader may be concerned that only the entropy
experienced corrections via the former method, whereas only the viscosity
was corrected by way of the latter. There is, however, no contradiction here.
Both the entropy and the viscosity for the Gauss-Bonnet theory will include
corrections from the scalar-squared and Ricci-squared terms.  The essential point
being that these particular corrections must be identical for any
choice of polarization and, hence, will  neatly cancel out of the entropy--viscosity
ratio. To put it another way,
while the entropy and viscosity can each vary (in unison) in going between the two theories, the ratio itself should only be sensitive to the Riemann-squared term. Reassuringly, this is precisely what we have found.


Let us next  consider how an arbitrary Lovelock theory will correct
the entropy--viscosity ratio. Perhaps counter-intuitively,
we will be able to demonstrate that the resulting correction is
rather simple, regardless of the dimensionality and/or order of the expansion.
Returning to the previous notations, we
can immediately deduce that $\mathscr{L}_{m-1}(g_{\parallel})=0$
is true for any $m$ (since all the applicable Riemann components
are vanishing on the horizon). Hence, the entropy coupling
$\kappa_{rt}^2$ will never vary from the Einstein value $\kappa^2_E$.

Now what about the quantity $\mathscr{L}_{m-1}(g_{\perp})$, which determines
the viscosity coupling $\kappa_{xy}^2$?
To address this query, it is helpful to first recall what is exactly meant by
the symbolism $g_{\perp}$. This notation instructs us to
incorporate only the curvature component $R^{rtrt}$ (and its
permutations) into the calculation. But, since the black brane metric
depends only on $r$, what we are  actually doing is calculating $\mathscr{L}_{m-1}$
for an effective two-dimensional theory; meaning that $\mathscr{L}_{m-1}(g_{\perp})$
must identically vanish for any $(m-1)>2/2=1$ or $m>2$. Hence,
any corrections to the viscosity coupling can {\em only} originate from
the Gauss-Bonnet term. Put differently, irrespective of the order of the Lovelock
expansion, absolutely nothing has changed from the preceding (Gauss-Bonnet)
calculation. Thus, we may conclude that eq.(\ref{etasGB}) with $\lambda=\lambda_2$
is true for {\em any} Lovelock theory of gravity.

\section{Acknowledgments}
We thank Dan Gorbonos for suggesting the problem to us and for several previous discussions with RB. We thank Alex Buchel for useful explanations and comments on the manuscript and Amos Yarom for comments on the manuscript.
The research of RB was supported by The Israel Science Foundation grant no 470/06.
The research of AJMM is supported by the University of Seoul.
AJMM sends many thanks to  Ben-Gurion University for their gracious hospitality during his visit.

\end{document}